\documentclass{aastex631}

	\usepackage{graphicx}	
	\usepackage{amsmath}
	\usepackage{amssymb}	
	\usepackage{longtable}
	\usepackage{threeparttable}
	\usepackage{multirow}
	\usepackage{booktabs}
	\usepackage{array}

	\shorttitle{AH Her}
	\shortauthors{Sun et al.}
	\graphicspath{{./}{figures/}}
	\received{April 5, 2023}
	\accepted{June 24, 2023}
		
\begin{document}
			
			\title{Evolution of negative superhumps, quasi-periodic oscillations and outbursts in the Z Cam-type dwarf nova AH Her}
		
\correspondingauthor{Sheng-Bang Qian, Qi-Bin Sun}					
\author{Qi-Bin Sun}\email{qsb@ynao.ac.cn, sunqibin@ynao.ac.cn}
\address{Yunnan Observatories, Chinese Academy of Sciences, Kunming 650216, People’s Republic of China}
\address{Department of Astronomy, Key Laboratory of Astroparticle Physics of Yunnan Province, Yunnan University, Kunming 650091, People’s Republic of China}
\affiliation{Center for Astronomical Mega-Science, Chinese Academy of Sciences, 20A Datun Road, Chaoyang District, Beijing, 100012, People’s Republic of China}
\affiliation{University of Chinese Academy of Sciences, No.19(A) Yuquan Road, Shijingshan District, Beijing, People’s Republic of China}

\author{Sheng-Bang Qian}
\affiliation{Yunnan Observatories, Chinese Academy of Sciences, Kunming 650216, People’s Republic of China}
\affiliation{Department of Astronomy, Key Laboratory of Astroparticle Physics of Yunnan Province, Yunnan University, Kunming 650091, People’s Republic of China}
\affiliation{Center for Astronomical Mega-Science, Chinese Academy of Sciences, 20A Datun Road, Chaoyang District, Beijing, 100012, People’s Republic of China}
\affiliation{University of Chinese Academy of Sciences, No.19(A) Yuquan Road, Shijingshan District, Beijing, People’s Republic of China}

\author{Min-Yu Li}
\affiliation{Yunnan Observatories, Chinese Academy of Sciences, Kunming 650216, People’s Republic of China}
\affiliation{Center for Astronomical Mega-Science, Chinese Academy of Sciences, 20A Datun Road, Chaoyang District, Beijing, 100012, People’s Republic of China}
\affiliation{University of Chinese Academy of Sciences, No.19(A) Yuquan Road, Shijingshan District, Beijing, People’s Republic of China}


			\begin{abstract}
	AH Her is a Z Cam-type dwarf nova with an orbital period of $ \sim $ 0.258 d. Dwarf nova oscillations and long-period dwarf nova oscillations have been detected, but no quasi-periodic oscillations (QPOs) and negative superhumps (NSHs) have been found. We investigated the association between NSHs, QPOs, and outbursts of AH Her based on \textit{TESS} photometry. We find for the first time the NSHs with a period of 0.24497(1) d in AH Her, and trace the variation of the amplitude and period of NSHs with the outburst. 
	The amplitude of the NSHs is most significant at quiescence, weakening as the outburst rises, becoming undetectable at the top, rebounding and weakening at the plateau, and strengthening again as the outburst declines. The variation of the accretion disk radius can explain the NSHs amplitude variation except for the plateau, so we suggest that the relationship between NSHs amplitude and outburst can be used as a window to study the accretion disk instability and the origin of NSHs. 
	In addition, we find the periodic variations in the amplitude, maxima, and shape of the NSHs ranging from 2.33(2) d to 2.68(5) d, which may be related to the precession of the tilted disk.
	Finally, we find QPOs at the top of AH Her's long outburst with $ \sim $ 2800 s similar to HS 2325+8205, suggesting that the presence of QPOs at the top of Z Cam's long outburst may be a general phenomenon.

\end{abstract}

\keywords{Binary stars; Cataclysmic variable stars; Dwarf novae; individual (AH Her)}


\section{Introduction} \label{sec:intro}

Dwarf novae (DNe) are a subtype of cataclysmic variables (CVs) with outburst amplitudes in the range of 2-8 magnitudes, lasting from several days to several weeks, and recurrence time from days to decades \citep{warner1995cat}. The outbursts of DNe can be explained by the thermal limit-cycle instability in the accretion disk (The disk instability model: DIM; \citealp{lasota2001disc}).
DNe can be divided into Z Camelopardalis stars, SU Ursae Majoris stars, and U Geminorum stars, three main subtypes \citep{warner1995cat}. The most characteristic feature of Z Cam is a brightness ``standstill" during the outburst recession stage, which is about 0.7 mag lower than the peak of the outburst \citep{lasota2001disc}. The DIM invokes a transition between the cold and hot states of the accretion disk. When the mass transfer from the secondary is larger than the critical rate ($\dot{M}_{crit}$), the accretion disk is in the hot state, as in the case of nova-like variables (NLs, \citealp{Smak1983ApJ}), another subtype of CVs.
Current studies show that the mass transfer rate of the Z Cam is close to $\dot{M}_{crit}$, and a slight change in the mass transfer rate can make a ``standstill" \citep{Meyer1983A&A,Smak1983ApJ,Honeycutt1998PASP}.
With the development of Z Cam research, it has been found that Z Cam have complex light curves. Two Z Cam-type DNe IW And and V513 Cas were first discovered to have the ``standstill" not end with returning to quiescence, but by an outburst followed by a deep dip \citep{2011JAVSO..39...66S}. \cite{2013PASP..125.1421S} subsequently confirmd this behavior, and called it an  ``anomalous standstill phenomenon".
During the ``Z CamPaign" observation campaign, the ``anomalous standstill phenomenon" was found in other Z Cam (e.g., HX Peg, AH Her, AT Cnc; \citealp{Simonsen2014JAVSO} ). \cite{Kato2019PASJ} named the objects with the ``anomalous standstill" as ``IW And-type objects (or phenomenon)." 

Negative superhumps (NSHs) are the light modulation of small amplitude with a period several percent shorter than the orbital period and are thought to come from the reverse precession of the nodal line of the tilted disk and the orbital motion of the binary star (e.g., \citealp{1985A&A...143..313B}; \citealp{patterson1999permanent}; \citealp{harvey1995superhumps}). 
The NSHs can be reproduced based on a tilted disk using the smoothed particle hydrodynamic (SPH) simulations (e.g., \citealp{wood2009sph}; \citealp{Montgomery2009MNRAS}; \citealp{Montgomery2012}). 
NSHs are being recognized in an increasing number of systems (e.g., \citealp{Olech2009MNRAS}; \citealp{2013MNRAS.435..707A};  \citealp{sun2022study,Sun2023arXiv}; \citealp{Bruch2022,Bruch2023}), as well as in low mass X-ray binaries (e.g., V1405 Aql; \citealp{retter2002detection} and GR Mus; \citealp{2013MNRASCornelisse}).

There are three types of rapid oscillations in CVs (see \cite{warner2004rapid} for details):

(i) Dwarf nova oscillation oscillations (DNOs);

(ii) Long-period dwarf nova oscillations (lp-DNOs);

(iii) Quasi-periodic oscillations (QPOs).

The periods of QPOs range from hundreds to thousands of seconds, and are longer than those of lp-DNOs and DNOs (QPOs $ \approx $ 4 $ \times $ lp-DNOs $ \approx $ 16 $ \times $ DNOs; \citealp{Pretorius2006MNRAS}).
\citet{warner2004rapid} proposed at least three types of QPOs, the first being DNO-related QPOs, which coexist with DNOs and oscillate for a few hundred seconds. The second type is IP-related QPOs, which oscillate with a period of about 1000 seconds and whose generation is related to the structure of the intermediate polars (IPs; \citealp{patterson2002superhumps}). Except for the DNO-related QPOs and the IP-related QPOs, the rest of the QPOs are of the third kind (Other QPOs;
\citealp{warner2004rapid}, \citealp{scepi2021qpos}).
The origin of QPOs remains inconclusive (e.g., \citealp{Patter1979Rapid}; \citealp{okuda1992disc};  \citealp{lubow1993wave}; \citealp{paczynski1978phenomenological}), but the consensus is that they exist in systems with high mass transfer rates, such as NLs and during DNe outbursts (such as HS 2325+8205; \citealp{Sun2023MNRAS}).

AH Her is a classical Z Cam-type DN and the ``standstill" has been observed many times (e.g., \citealp{Honeycutt1998PASP}; \citealp{Spogli2011}; \citealp{Waagen2013AAN}), in addition to the anomalous brightness ``standstill" (e.g., \citealp{Simonsen2014JAVSO}; \citealp{Spogli2021}; \citealp{Ohshima2023arXiv}), so it is classified as an IW And-type object.
It was among the first DNe found to have DNOs \citep{Warner1972}, with an oscillation period of 24.0–38.8 s (e.g., \citealp{Patterson1978ApJ}; \citealp{Nevo1978MNRAS}; \citealp{Hildebrand1981ApJ}), and was also found to have an lp-DNO of about 100 s \citep{Patterson1981ApJS}, but no QPOs have been found.

\cite{Horne1986MNRAS} determined the period to P$_{\rm orb} $ = 0.258116 $ \pm $ 0.000004 d based on spectroscopic observations, and the other system parameters were determined as $ i $ = 46 $ \pm $ 3$ ^{\circ} $, $ q $ = 0.80 $ \pm $ 0.05, $  M_{1} $ = 0.95 $ \pm $ 0.10 $ M_{\bigodot} $ and $  M_{2} $ = 0.76 ± 0.08 $M_{\bigodot} $.
Based on the spectroscopic observations \cite{Bruch1987A&A} estimated that the spectral type of the secondary star might be K5.
\cite{Echevarr2021} performed photometric and spectroscopic observations combined with the radial velocities analysis for AH Her, and the orbital period was determined as P$ _{\rm orb}  $ = 0.25812 ± 0.00032 d in general agreement with \cite{Horne1986MNRAS}, but \cite{Horne1986MNRAS}'s orbital period is more accurate.

In this paper, NSHs and QPOs were first discovered in AH Her, and the primary purpose is to contribute to the connection and transformation between NSHs, QPOs, and DNe outbursts. This paper is structured as follows.
Section \ref{sec:style} presents the \textit{TESS} photometry of AH Her.
In Section \ref{sec:Analysis}, the NSHs and orbital period are determined.
Section \ref{sec:Changes} analyzes the NSHs amplitude variation.
In Section \ref{sec:O-C}, the O-C of maxima and slope variation of NSHs are analyzed.
Section \ref{sec:QPOs}, the QPOs are analyzed.
The NSHs amplitude, O-C, and slope variation are discussed in Section \ref{sec:DISCUSSION}, respectively.
Section \ref{sec:CONCLUSIONS} is the summary.


\section{\textit{TESS} Photometry } \label{sec:style}

This work used data from the \textit{Transiting Exoplanet Survey Satellite} (\textit{TESS}, \citealp{Ricker2015journal}), and which were generated by the \textit{TESS} Science Processing Operations Center (SPOC, \citealp{Jenkins2016}) of NASA Ames Research Center.
Photometry was performed in the wavelength range of 600 to 1000 nm \citep{Ricker2015journal}. The sky was divided into 26 overlapping sectors, each of which was observed for about one month. Observations were either long-cadence (30 min) or short-cadence (2 min) mode.

\begin{figure}
\centering
\includegraphics[width=0.7\columnwidth]{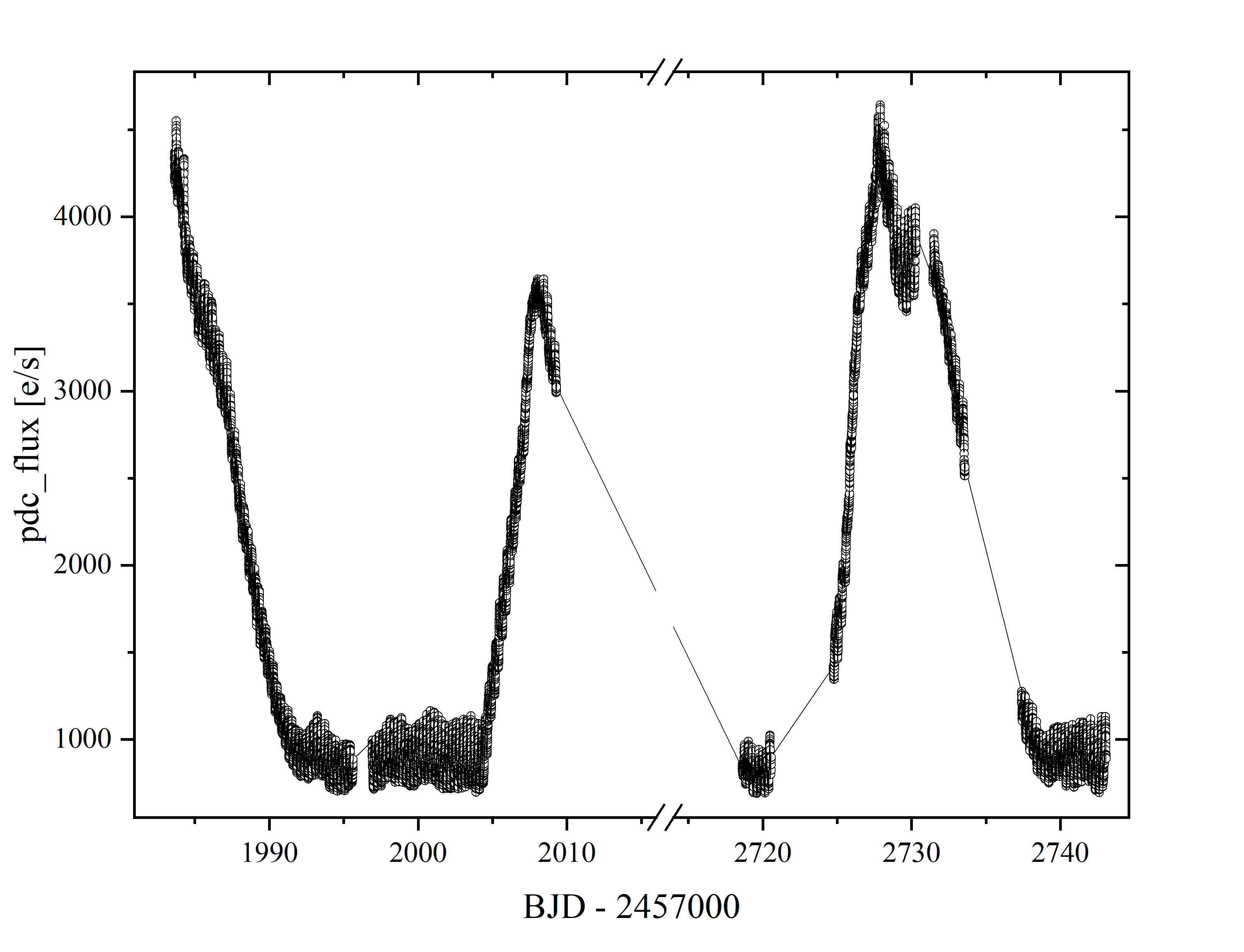}
\caption{The outburst of AH Her observed by \textit{TESS}.\label{fig:OUT}}
\end{figure}

AH Her was observed by \textit{TESS} photometry in sectors 25 (s25) and 52 (s52), respectively. The observation periods ranged from MBJD58983.13533 (2020-05-14) to 59008.80891 (2020-06-08) and MBJD59718.13786 (2022-05-19) to 59742.57939 (2022-06-12), respectively, with the short-cadence mode. 
Data were downloaded from the Mikulski Archive for Space Telescopes (MAST)\footnote{https://mast.stsci.edu/}.  
In our work, the PDCSAP\_FLUX (see \citealp{2010SPIE.7740E..1UT} for details) data are used (see Fig. \ref{fig:OUT}).

\begin{table*}
\centering
\caption{Results of the analysis of \texttt{Period04}.\label{tab:freq}}
\setlength{\tabcolsep}{2mm}{
	\begin{tabular}{llllllllll} 
		\toprule
		Sectors&  Name &  frequency & errors&  period& errors&  amplitude& errors& phase& errors\\
		&  & [1/d] & [1/d]& [d]& [d] & [e/s or e/(s$ \cdot $d)]  & [e/s or e/(s$ \cdot $d)]  & &   \\
		\midrule
		
		s25&Amplitude$^{\rm a}$&0.4152&0.0092&2.4088&0.0532&23.96&9.58&0.509&0.064\\
		&O-C&0.3945&0.0058&2.5351&0.0371&0.0123&0.0032&0.621&0.041\\
		&S$ _{\rm r}$$^{\rm b}$&0.3812&0.0032&2.6235&0.0220&419.61&59.06&0.684&0.022\\
		&S$ _{\rm d} $$^{\rm b}$&0.4016&0.0024&2.4903&0.0149&602.97&63.56&0.548&0.017\\
		&NSHs&4.07683&0.00016&0.245288&0.000010&89.00&0.66&0.147&0.001\\
		&2$ \times $f$ _{\rm orb}$$^{\rm c}$&7.74513&0.00058&0.129113&0.000010&24.45&0.66&0.528&0.004\\
		&QPOs$^{\rm d}$&30.9524&0.0366&0.032308&0.000038&36.72&2.42&0.452&0.010\\				
		s52&O-C &0.3736&0.0074&2.6769&0.0533&0.0125&0.0041&0.519&0.052\\
		&S$ _{\rm r}$$^{\rm b}$&0.4288&0.0039&2.3319&0.0214&408.55&70.70&0.560&0.028\\
		&S$ _{\rm d}$$^{\rm b}$&0.3753&0.0036&2.6646&0.0253&654.85&101.47&0.806&0.025\\
		&NSHs&4.08747&0.00034&0.244650&0.000020&71.95&1.08&0.307&0.002\\
		&2$ \times $f$ _{\rm orb}$$^{\rm c}$&7.74430&0.00112&0.129127&0.000019&21.67&1.08&0.247&0.008\\
		
		&QPOs$^{\rm d}$&31.4816&0.0312&0.031765&0.000032&26.64&1.99&0.750&0.012\\

		\bottomrule
\end{tabular}}
\begin{threeparttable} 
	\begin{tablenotes} 
		\item [a] NSHs amplitude variation parameters in s25 (see Section \ref{sec:Changes} for details). 
		\item [b] S$ _{\rm r}$ and S$ _{\rm d}$ represent the rising and declining slopes, respectively, and the amplitude corresponds to the unit of [ e/(s$ \cdot $d)] with apparent periodicities (see Section \ref{sec:O-C} for details). 
		\item [c] Two times the orbital frequency. 
		\item [d] QPOs at the top of the long outburst. 
	\end{tablenotes} 
\end{threeparttable}
\end{table*}


\section{Determination of NSHs and orbital period} \label{sec:Analysis}

\begin{figure}
\centering
\includegraphics[width=0.9\columnwidth]{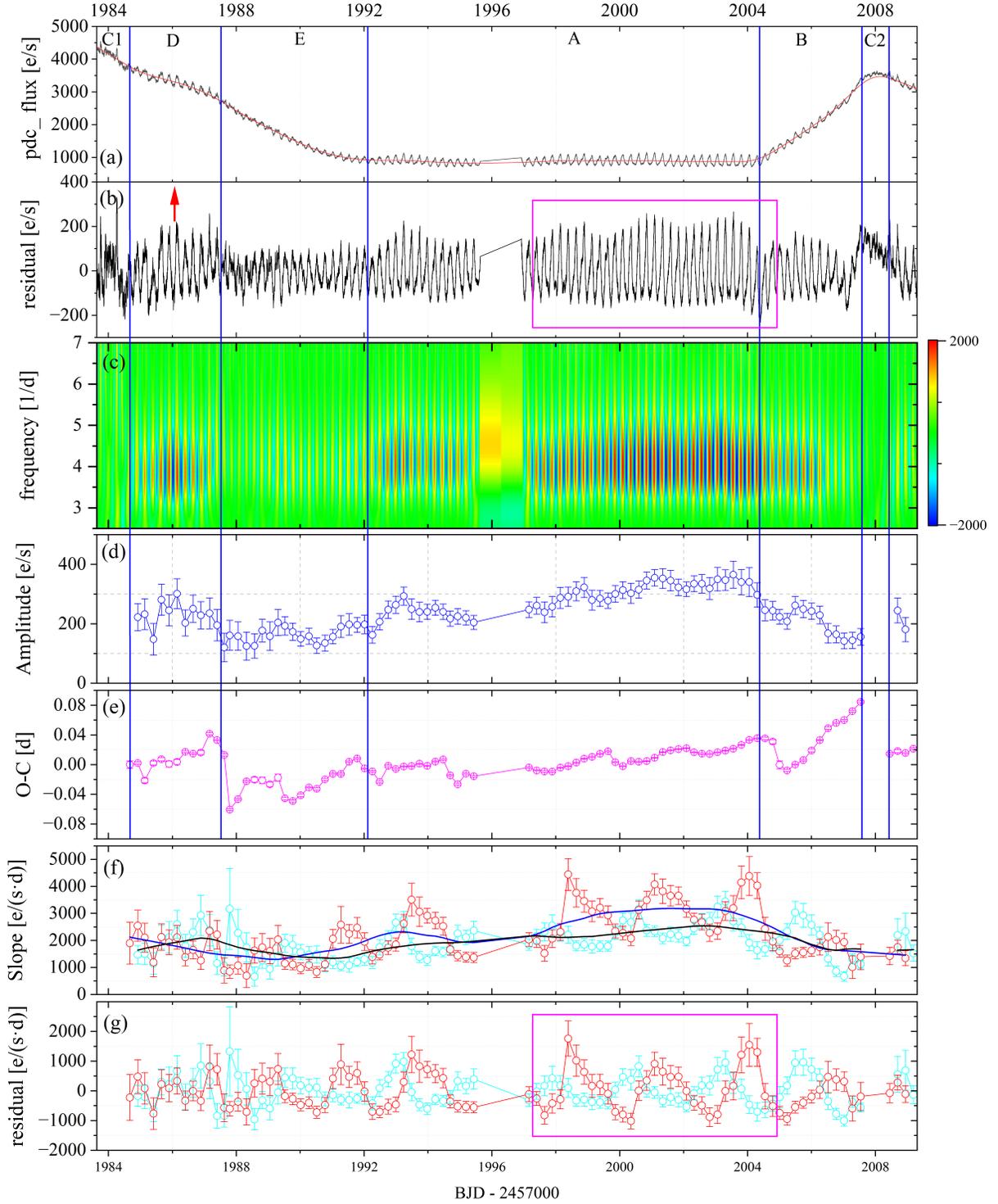}
\caption{The results of the analysis of the s25 light curve. (a): Two incomplete outbursts, the solid red line is the LOESS fit, and A-E represents the five different outburst and NSHs stages (see Section \ref{sec:Changes} for details); (b): Residuals of the LOESS fit; (c) CWT 2D spectrum of the light curve of the de-outburst trend; (d): Amplitude variation curve of the NSHs; (e): O-C curve of the NSHs maxima; (f): The red circle and cyan color are the declining and rising slopes of NSHs respectively, the blue and black solid lines are the LOESS fit; (g): The residuals of the LOESS fit in plate f.\label{fig:s25}}
\end{figure}

\begin{figure}
\centering
\includegraphics[width=0.9\columnwidth]{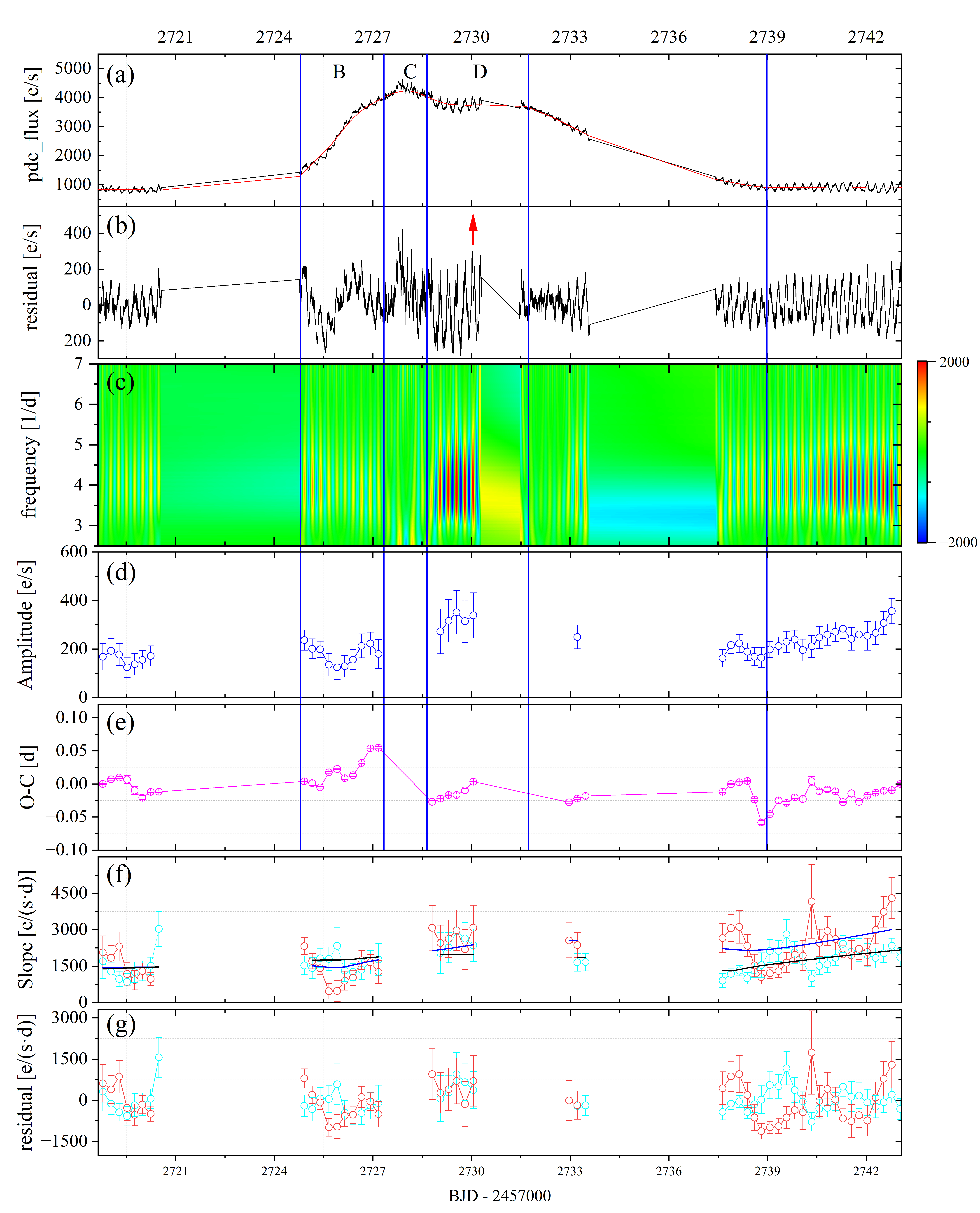}
\caption{The results of the analysis of the s52 light curve. The interpretation of each plate is consistent with Fig. \ref{fig:s25}, except that the light curve is replaced with s52.\label{fig:s52}}
\end{figure}

\begin{figure}
\centering
\includegraphics[width=0.7\columnwidth]{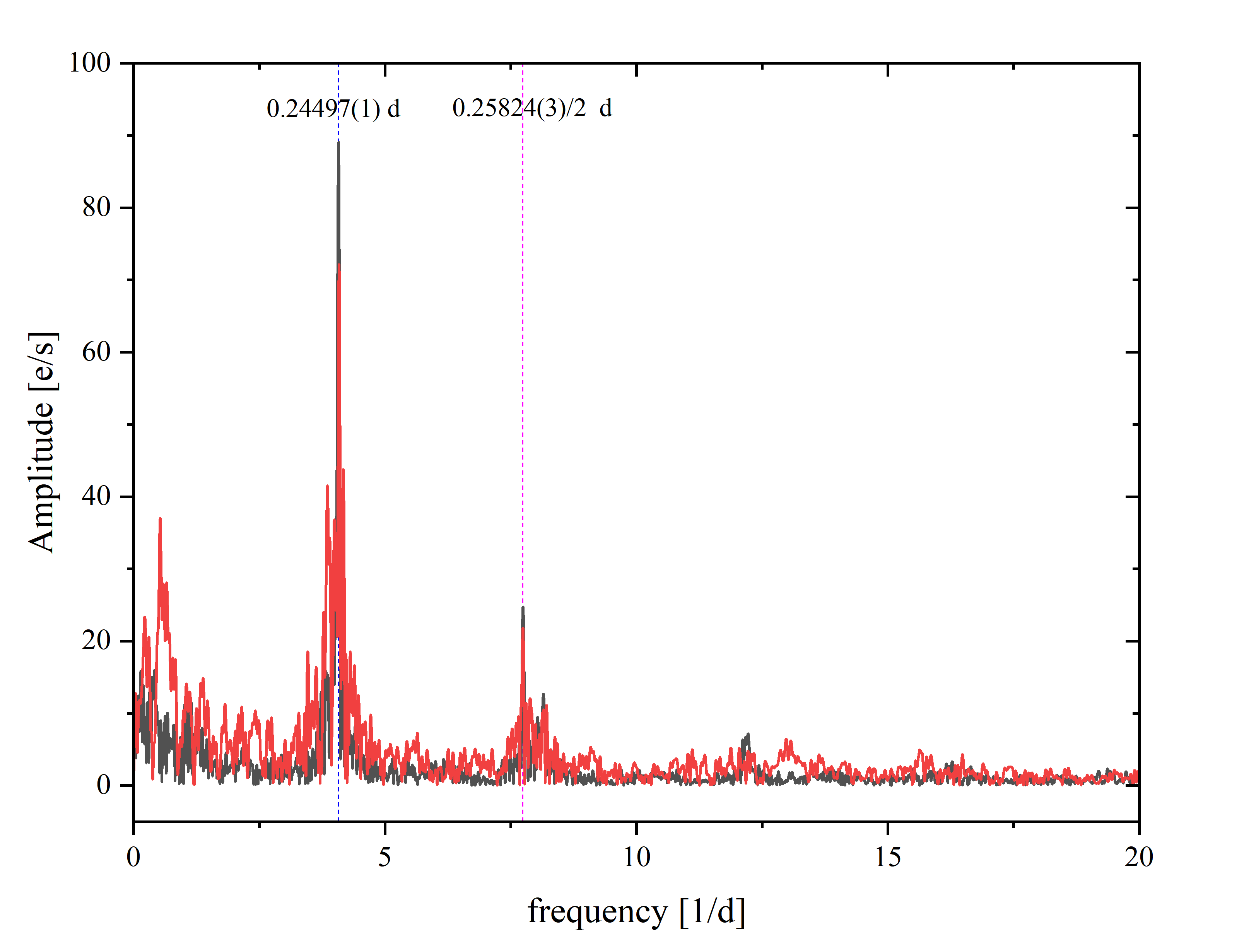}
\caption{Frequency-Amplitude spectrum of AH Her. The black and red solid lines correspond to the analysis results of s25 and s52, and the blue and magenta dashed vertical lines correspond to 0.24497(1) d and 0.25824(3) / 2 d .\label{fig:p}}
\end{figure}

The \textit{TESS} photometry has three incomplete outbursts (see Fig. \ref{fig:OUT}).
The outburst trend was first removed using the locally weighted regression (LOESS; \citealp{cleveland1979robust}) fit spanning 0.05 d (see Fig. \ref{fig:s25} a and \ref{fig:s52} a). Then the light curves with outburst trends removed in s25 and s52 were searched for periods using \texttt{Period04} (see \cite{Period04} for details), respectively.
The analysis show signals of $ \sim $ 0.245 d and $ \sim $ 0.129 d in both s25 and s52 (see Table \ref{tab:freq} and Fig. \ref{fig:p}).
Combined with the orbital period determined by \cite{Horne1986MNRAS}, it can be determined that the 0.245 d period is $\sim$ 5 per cent shorter than the 0.258116(4) d orbital period, and appears to be an NSH. The 0.129 d period appears to be a second harmonic of $P_{\rm orb}$, that is double the orbital frequency.
The average of the results of the two sectors is taken, and the orbital period of AH Her was determined to be P$_{\rm orb}  $ = 0.25824(3) d. In addition, the period of NSHs averaged over the two sectors was determined to be 0.24497(1) d, and the excess ($\epsilon ^- = (P_{\rm orb} - P_{\rm nsh})/P_{\rm orb}$ ) is calculated as $\epsilon ^- \sim 0.051$.

The orbital period of AH Her was accurately determined by \cite{Horne1986MNRAS} to be P$_{\rm orb}  $ = 0.258116(4) d based on the radial velocity analysis (spectroscopic observations in 1980 and 1981). \cite{Echevarr2021} also performed the radial velocity analysis (spectroscopic observations in 2013) and obtained an orbital period of P$_{\rm orb}  $ = 0.25812(32) d. The time interval between the two spectroscopic observations is about 30 years but the difference in the orbital period is within the uncertainty, which proves that the orbital period of AH Her is relatively stable. 
The accuracy of the orbital period we obtained (error = 3 $\times$ 10$^{-5}$ d) is less than \cite{Horne1986MNRAS} (error = 4 $\times$ 10$^{-6}$ d) but greater than \cite{Echevarr2021} (error = 3.2 $\times$ 10$^{-4}$ d). The differences between our determined orbital period and theirs are all within the uncertainty of \cite{Echevarr2021}. In addition, the orbital period we obtained is based on a large number of photometric data from \textit{TESS} and have an important reference value.

\section{Changes in NSHs amplitude} \label{sec:Changes}

To investigate the relationship between NSHs and outbursts more comprehensively, we used the Continuous Wavelet Transform (CWT; \citealp{polikar1996wavelet}) to analyze the time-domain information of NSHs. The data from both sectors were analyzed separately, as above. Using CWT for the light curve with the outburst trend removed, the results show that NSHs decay with the rise of the outburst and are strongest in the quiescence.
In addition, we calculate the amplitude of each individual cycle of the NSH modulation combined with the CWT results to investigate the evolution of the NSHs with the outburst. The calculation of the amplitude is based on the maxima and minima of the NSHs.
Firstly, the linear superposition Gaussian fit was used to calculate the maxima and minima of the NSHs, with errors coming from the covariance (see Fig. \ref{fig:example} b). 142 maxima and 142 minima were obtained (see Supplementary Material). 
Secondly, based on the maxima and minima, the amplitude of the NSHs comes from the difference between the average of the two minima before and after the maxima and the maxima ($ \vartriangle \rm Amplitude = (\rm maxima $ $_{\rm flux} - (\rm minima $ $_{\rm before, \rm flux} + \rm minima $ $_{\rm after, \rm flux}) / 2  $). The results are shown in Figs. \ref{fig:s25} d and \ref{fig:s52} d. 
Combining the light curves, CWT 2D spectrum, and the NSHs amplitude, a clear correlation between NSHs and the outburst can be found. The NSHs can be divided into five states (Figs. \ref{fig:s25} and \ref{fig:s52}): 

Quiescence (State A): The maximum amplitude occurs during the quiescence. 

Rise (State B): During the outburst rise, the NSHs gradually weakened. 

Top (State C): At the top of the outburst, NSHs become undetectable.

Plateau (State D): Recurrence of NSHs in the outburst plateau, and the intensity of NSHs first enhanced and then weakened.

Decline (State E): During the decline of the outburst, the NSHs shift from a recurrence in the plateau to a weakening and then strengthen again as it gradually returns to quiescence.

This series of changes can be seen by the AH Her light curve, CWT 2D spectrum, and the change of NSHs amplitude.
Plateau (Stage D) is extraordinary among the five stages, and we use the upward arrows to label the rebound of NSHs. 
It can be seen that there is no significant rebound in the amplitude of NSHs during the outburst rise (State B), but there is a significant rebound from the peak of the outburst down to the quiescence (State A) in the plateau (State D), which implies that there is a difference in the physics scenario of the outburst rise and decline.
The NSHs evolve from the top of the outburst (where NSHs are undetectable) to a plateau that first strengthens and then weakens, and then strengthens again as the outburst declines. The critical finding of an apparent rebound of NSHs in the plateau provides a new basis for studying DNe outbursts and NSHs.

In addition to the variation of NSHs amplitude with the outburst, a periodic variation can be found in s25 (Fig. \ref{fig:s25} d). 
The analysis of the amplitude variation curve identifies the period as 2.41(5) d (see Fig. \ref{fig:poc}), but it is not significant in s52 due to the lack of data. We will use a sinusoidal fit for verification.   
The trend in amplitude with outburst is first removed using LOESS fit with a span of 0.2 d, and then the residuals are fitted using a least-squares sinusoidal model (see Fig. \ref{fig:amp}):
\begin{equation}\label{eq:sine}
{\rm flux(t)} = \rm{flux_0} + A \sin(2 \pi (t - T_0) / P)
\end{equation}
$\rm flux_0$, $\rm A$, $\rm T_0$, and $\rm P$ are the fitted constant, amplitude, phase, and period, respectively. The error in the fitted parameters comes from the covariance. The best-fit results are shown in Table \ref{tab:sine}. the best-fit period is 2.42(2) d in agreement with the results of the period analysis. The periodic variations in NSHs amplitude can be found by comparing amplitude variation curves with the light curve (see Figs. \ref{fig:s25} b, \ref{fig:example} b and c). The periodic variation of the NSHs amplitude provides a window to study the origin of the NSHs, which will be discussed in Section \ref{sec:disk}.

\begin{figure}
\centering
\includegraphics[width=0.7\columnwidth]{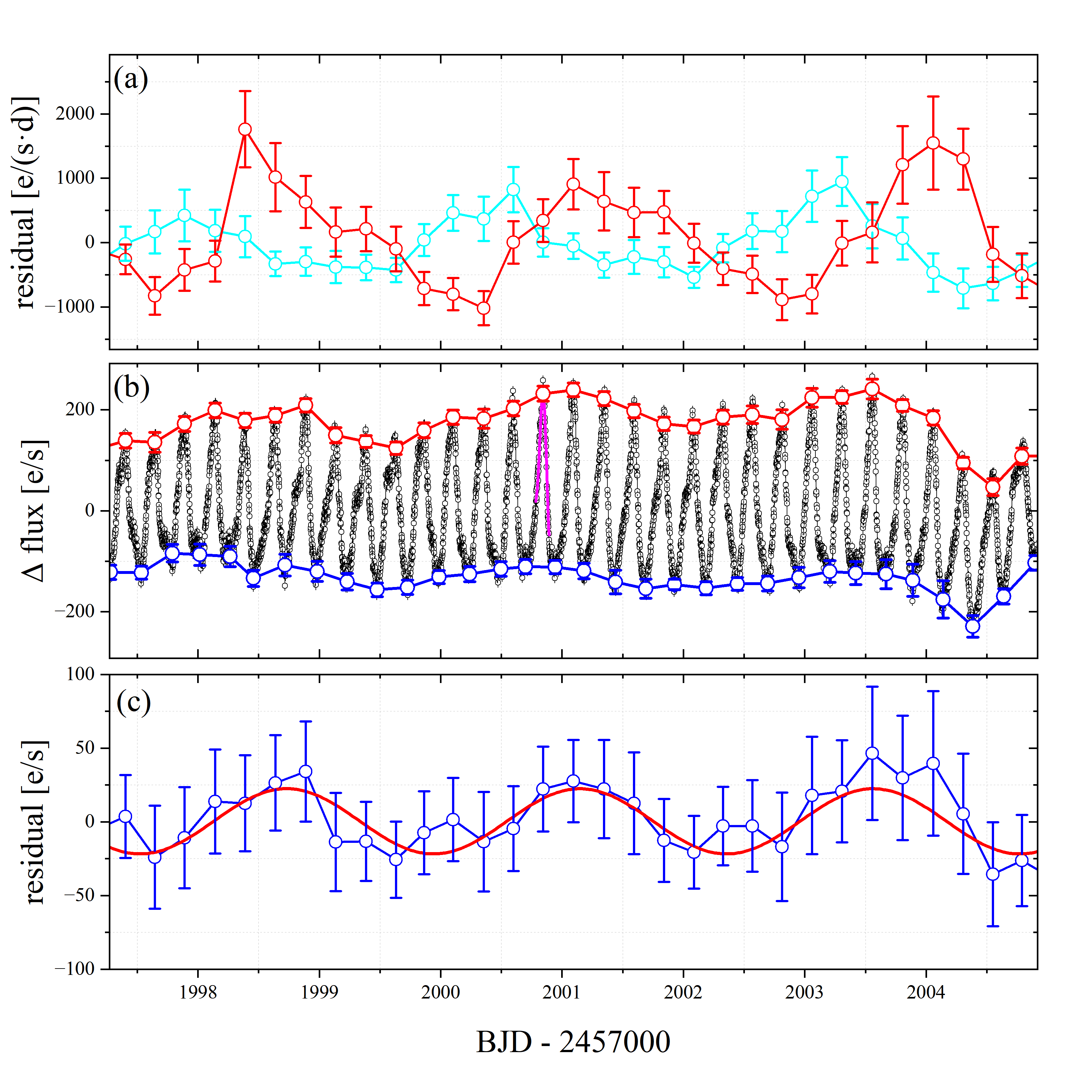}
\caption{Example of calculating the maxima and minima of NSHs and an example of slope and amplitude variations. Top panel: Example of slope variation from the magenta rectangular box Fig. \ref{fig:s25} g. Middle panel: the red and blue circles represent the maxima and minima, respectively, and the solid magenta line is a linear superposition Gaussian fit with a fitted width of 0.5 NSHs phase; light curve from the magenta rectangular box in Fig. \ref{fig:s25} b; Bottom panel: partial amplitude variation and sine fit from the magenta rectangular box in Fig. \ref{fig:amp}.\label{fig:example}}
\end{figure}

\begin{figure}
\centering
\includegraphics[width=0.7\columnwidth]{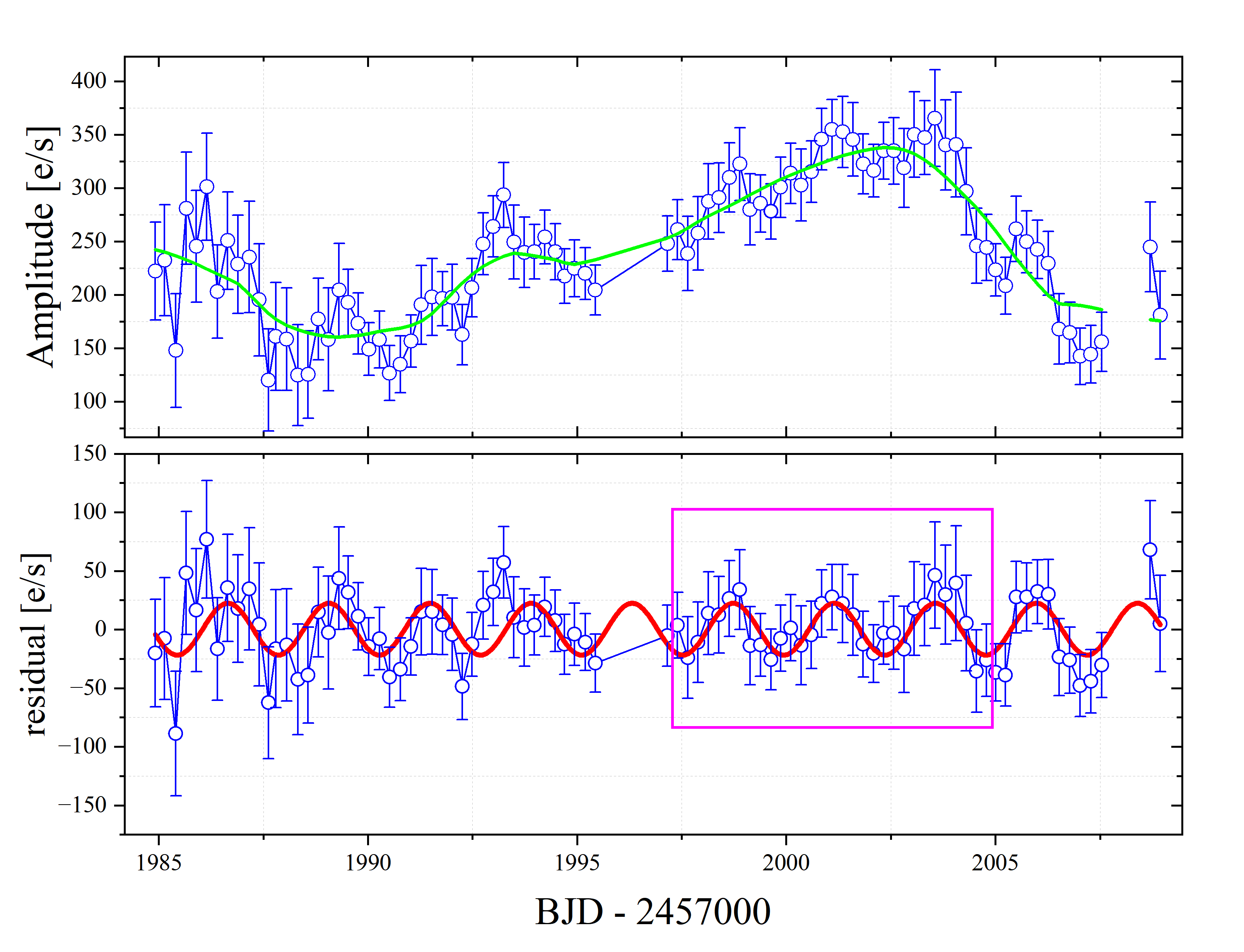}
\caption{Validation plot of the amplitude periodic variation of NSHs. The top panel is consistent with Fig. \ref{fig:s25} d, and the solid green line is the LOESS fit; the bottom panel is the residuals of the LOESS fit, and the solid red line is the sine fit.\label{fig:amp}}
\end{figure}

\begin{table*}
\centering
\caption{Results of sinusoidal fitting of NSHs amplitudes and QPOs.\label{tab:sine}}
\setlength{\tabcolsep}{1.5mm}{
	\begin{tabular}{lllllllllllll} 
		\toprule
		Name&Start$^{\rm a}$ & End$^{\rm a}$ &  flux$_{0} $&Errors&A&Errors&P&Errors&T$ _{0} $&Errors & $\nu$$^{\rm b}$ &$\chi^2_{red}$$^{\rm c}$\\
		& [d]& [d] & [e/s]&[e/s]&[e/s]&[e/s]&[d]&[d]& &  & &\\
		\midrule
		Amplitude&1984.67147&2009.18977&0.35&2.35&12.09&17.86&2.416&0.022&22.10&3.27&85&0.47\\			
		
		QPOs(s25)&1983.63683&1984.62711&-0.31&1.71&44.74&2.30&0.03231&3.84E-05&36.82&2.41&703&15.90\\
		QPOs(s52)&2727.44909&2728.76993&-0.15&1.41&-25.09&2.67&0.03177&3.08E-05&27.10&1.99&948&19.57\\

		\bottomrule
\end{tabular}}
\begin{threeparttable} 
	\begin{tablenotes} 
		\item [a] BJD2457000+. 
		\item [b] Degree of freedom of the fit. 
		\item [c] Reduced chi-square.
	\end{tablenotes} 
\end{threeparttable}
\end{table*}

\section{O-C analysis and shape change} \label{sec:O-C}

\begin{figure}
\centering
\includegraphics[width=0.7\columnwidth]{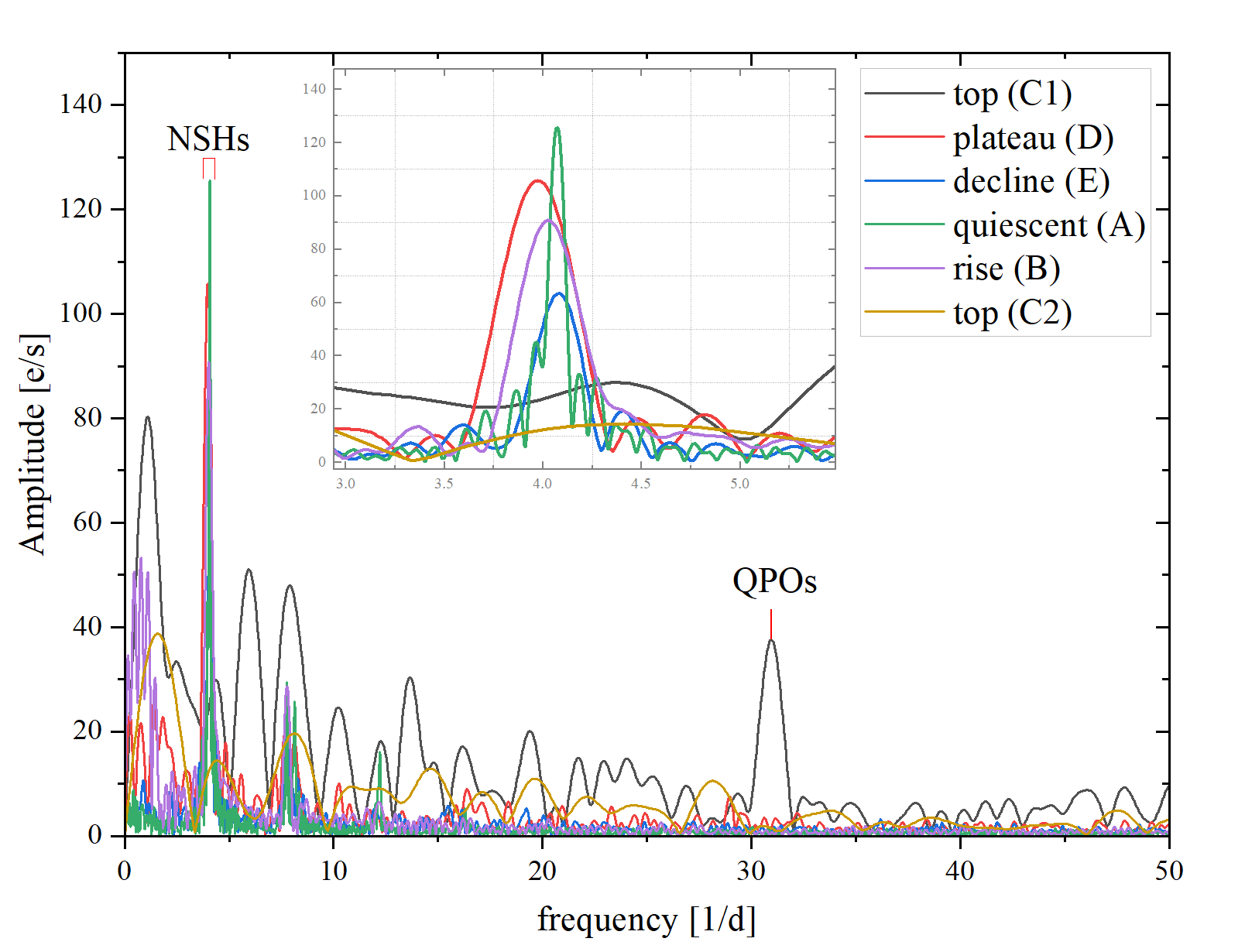}
\caption{Frequency-Amplitude spectrum the s25 light curve in different stages. The meanings of the different curves are shown in the legend, and the letters correspond to the individual stages in Fig. \ref{fig:s25} a.\label{fig:abc}}
\end{figure}

In Section \ref{sec:Changes}, we found that the amplitude of the NSHs varies with the outburst, in addition to the periodicity, with a period of 2.41(5) d.
To explore the variation of NSHs in more depth, this subsection will analyze the O-C of the maxima and the variation of the shape.
Based on the linear superposition Gaussian fit, 142 maxima were obtained, including 90 in s25 and 52 in s52. The following two ephemerides were obtained using the first maximum of each of the two sectors as the initial epoch and 0.24497(1) as the period:

\begin{equation} \label{eq:ephemeris1}
\mathrm{Maximum}(\rm s25) = \mathrm{BJD2458984.6715}(52) + 0.24497(1) \times \rm E 
\end{equation}
\begin{equation} \label{eq:ephemeris2}
\mathrm{Maximum}(\rm s52) = \mathrm{BJD2459718.7841}(9) + 0.24497(1) \times \rm E 
\end{equation}
The O-C analysis was performed using these two ephemerides.
By comparing the outburst trend with the O-C variation, it can be found that the O-C increases when the outburst rises in both sectors (see Figs. \ref{fig:s25} c and \ref{fig:s52} c), indicating that the period of NSHs increases as the outburst rises. For additional validation, we detrend the data of \textit{TESS} s25 (data relatively complete) for frequency analysis by different outburst phases. The results further indicate that the NSHs in the quiescence have the shortest period with the largest amplitude, and the NSHs disappear at the top of the outburst (see Fig. \ref{fig:abc}), which is consistent with the CWT 2D spectrum results.

\begin{figure}
\centering
\includegraphics[width=\columnwidth]{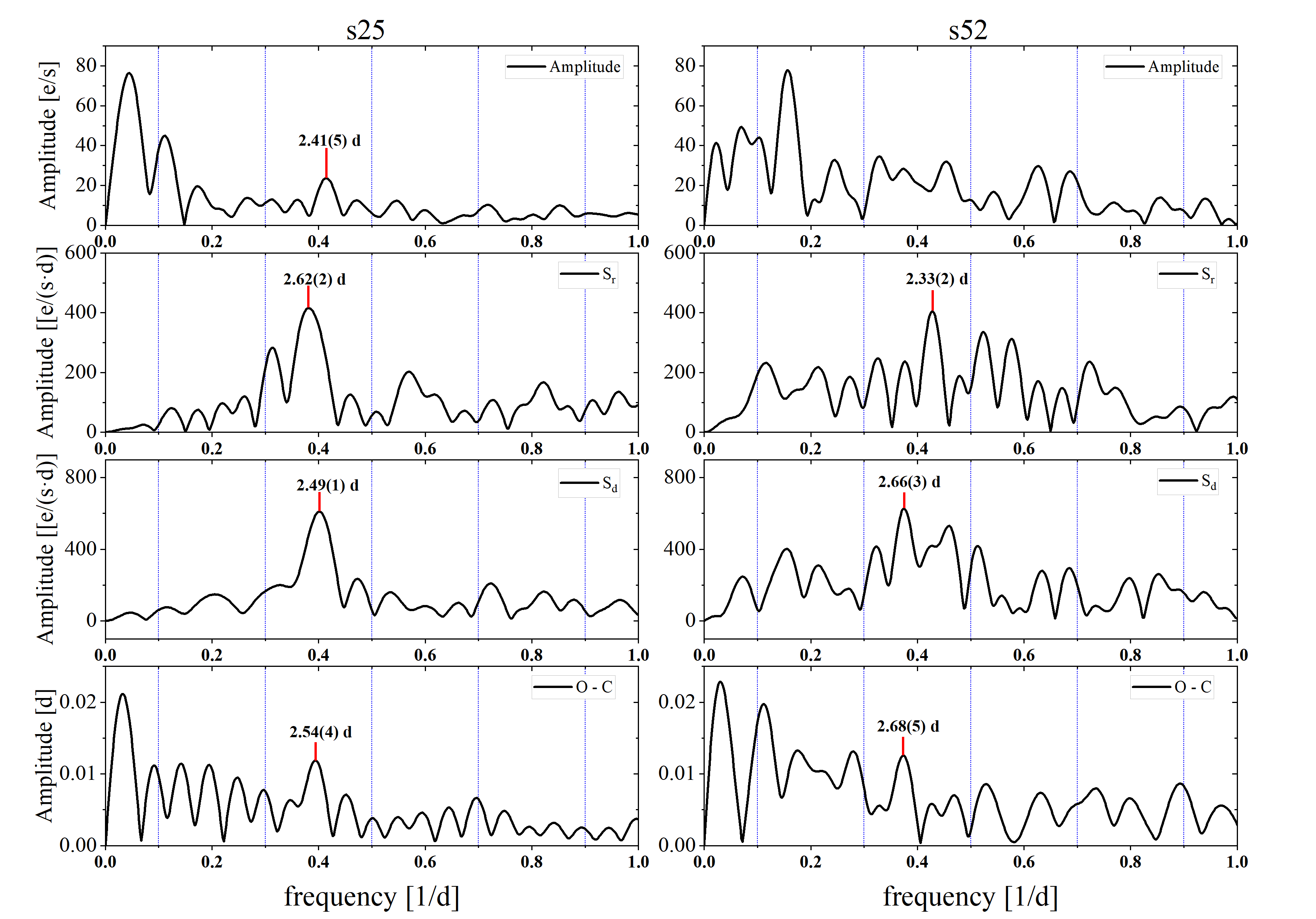}
\caption{Frequency-Amplitude plots of NSHs amplitude, maxima O-C, and slope variation. The four plots on the left are from the analysis of the \textit{TESS} s25 data, and the right is from s52. From top to bottom, NSHs amplitude, rising slope, declining slope, and O-C period analysis are shown.\label{fig:poc}}
\end{figure}

In addition to this, there may be periodic variations in O-C. The results of the analysis using \texttt{Period04} show possible periodic variations in O-C of 2.54(4) d and 2.68(5) d for s25 and s52, respectively, but they are not significant due to the lack of data (see the bottom panel in Fig. \ref{fig:poc}).
Because periodicities were not significant in the O-C data, we explored the Slope for each NSH rise and decline to quantify the analysis. 
The rising (S$ _{\rm r} $) and declining (S$ _{\rm d} $) slopes of NSHs come from the calculation between the minima and maxima  ($\rm S = {\Delta {\rm flux} \over \delta t}$), and the S$ _{\rm d} $ is taken as positive for the convenience of comparison. The results show that the Slope of the quiescence is larger than the outburst because of the large NSHs amplitude. In addition, an apparent periodic variation of both S$ _{\rm r} $ and S$ _{\rm d} $ can be found (see Figs. \ref{fig:s25} f and \ref{fig:s52} f). To detect this period, we used LOESS fits spanning 0.2 d and 0.5 d to remove trends in slopes (both S$ _{\rm r} $ and S$ _{\rm d} $) for s25 and s52, respectively, and finally analyzed the detrended data (see Figs. \ref{fig:s25} g and \ref{fig:s52} g). The results of the period analysis show that there are periodic variations of 2.62(2) d and 2.49(1) d in the S$ _{\rm r}$ and S$ _{\rm d} $ in s25, and periods of 2.33(2) d and 2.66(3) d in the S$ _{\rm r}$ and S$ _{\rm d} $ in s52, respectively (see Fig. \ref{fig:poc} and Table \ref{tab:freq}). In addition, to verify the variation of the Slope, we selected some curves of the quiescence of s25 to compare with the variation of the Slope, and the results further verified the existence of periodic variation of the shape of NSHs (see Fig. \ref{fig:example} a and b).

Based on the maxima O-C analysis of the NSHs and the change in Slope, the results altogether point to the presence of a 2.33(2) d to 2.68(5) d change in the NSHs (see Fig. \ref{fig:poc} and Table \ref{tab:freq}), which provides a new window for studying the NSHs of AH Her.

\section{Quasi-periodic oscillations} \label{sec:QPOs}

\begin{figure}
\centering
\includegraphics[width=0.65\columnwidth]{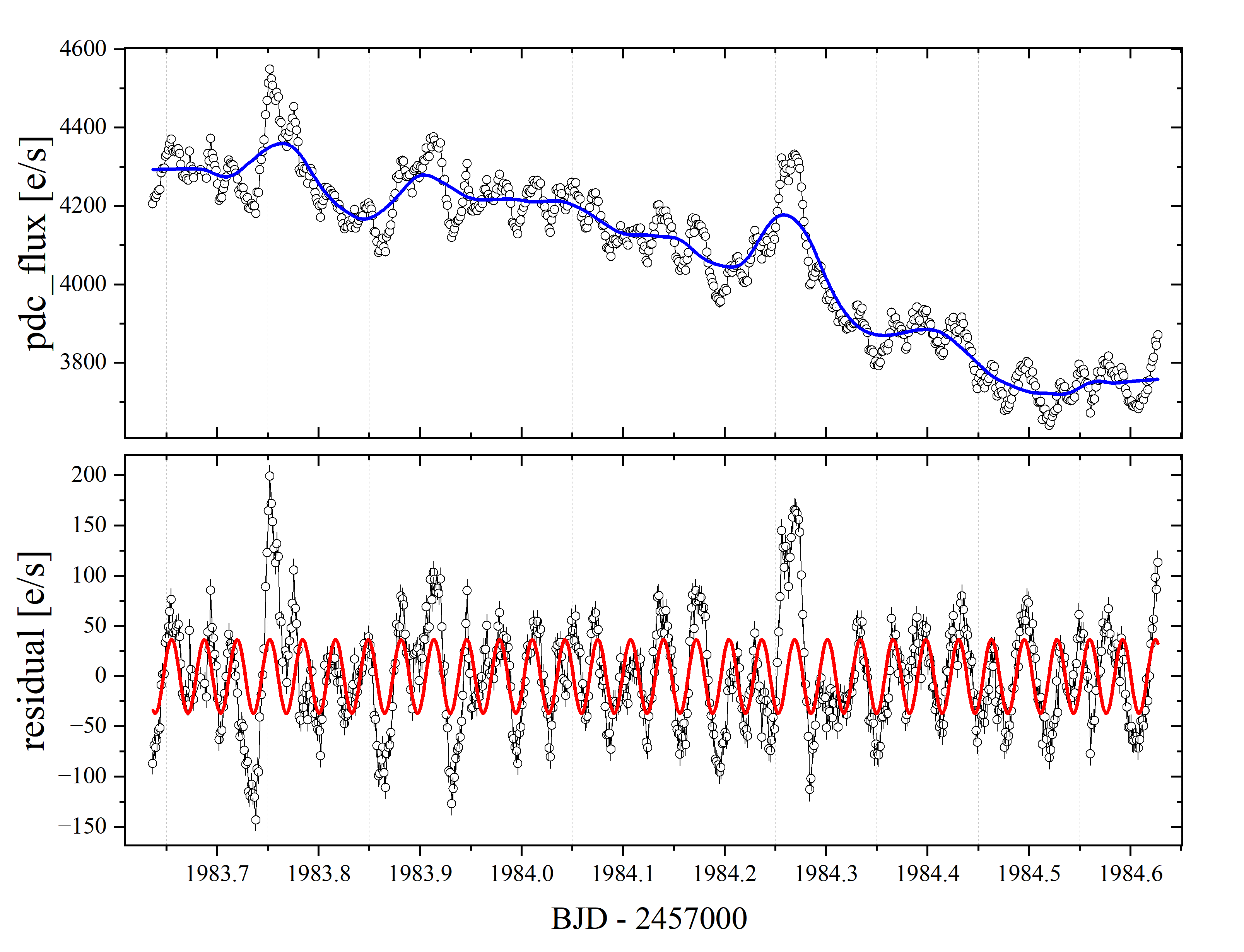}
\caption{Analysis of QPOs at the top of long outbursts in \textit{TESS} s25. The solid blue line in the top panel is the LOESS fit; the bottom panel is the residuals of the LOESS fit, and the solid red line is the sine fit. \label{fig:QPO1}}
\end{figure}

\begin{figure}
\centering
\includegraphics[width=0.65\columnwidth]{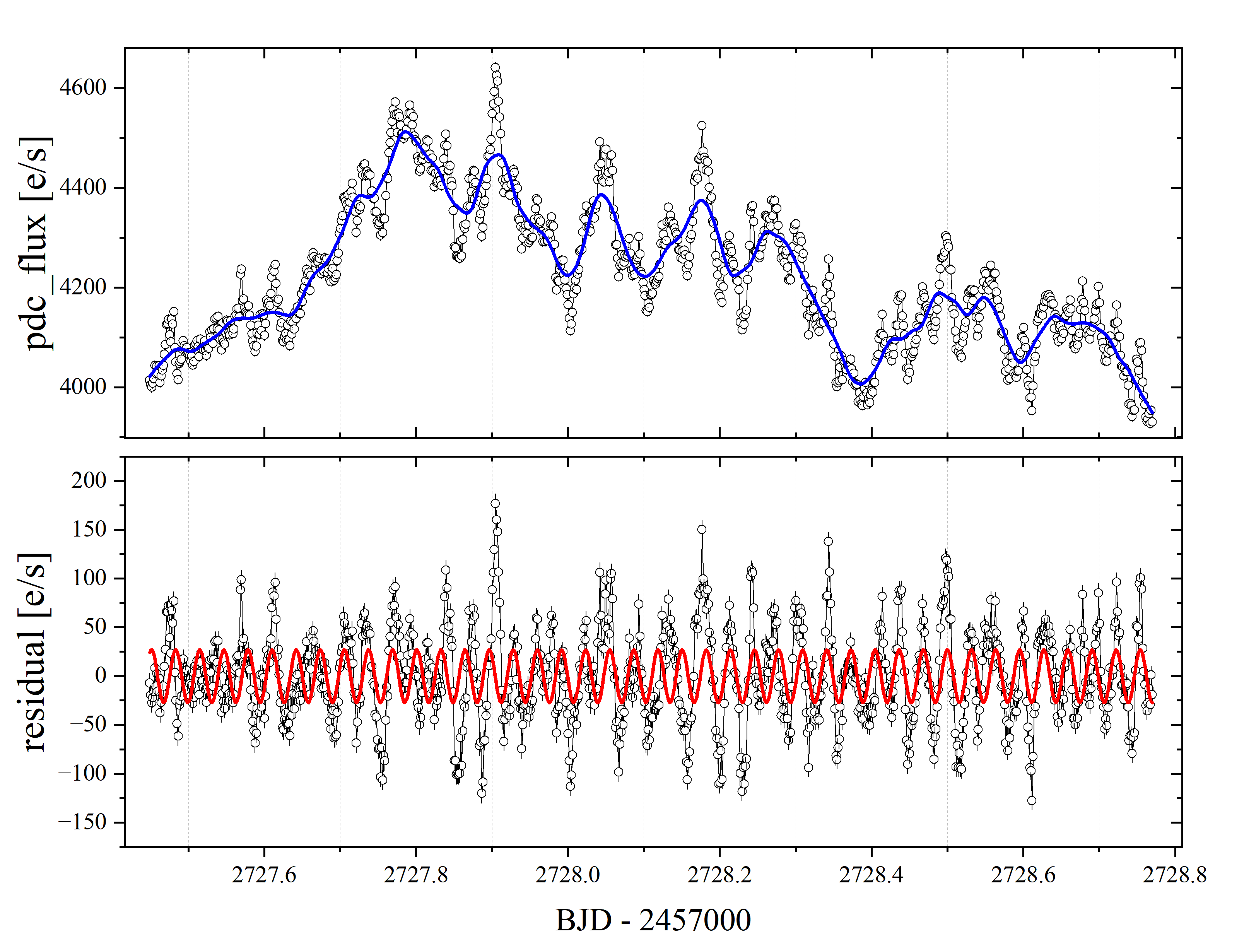}
\caption{Analysis of QPOs at the top of outbursts in \textit{TESS} s52. \label{fig:QPO2}}
\end{figure}

\begin{figure}
\centering
\includegraphics[width=0.7\columnwidth]{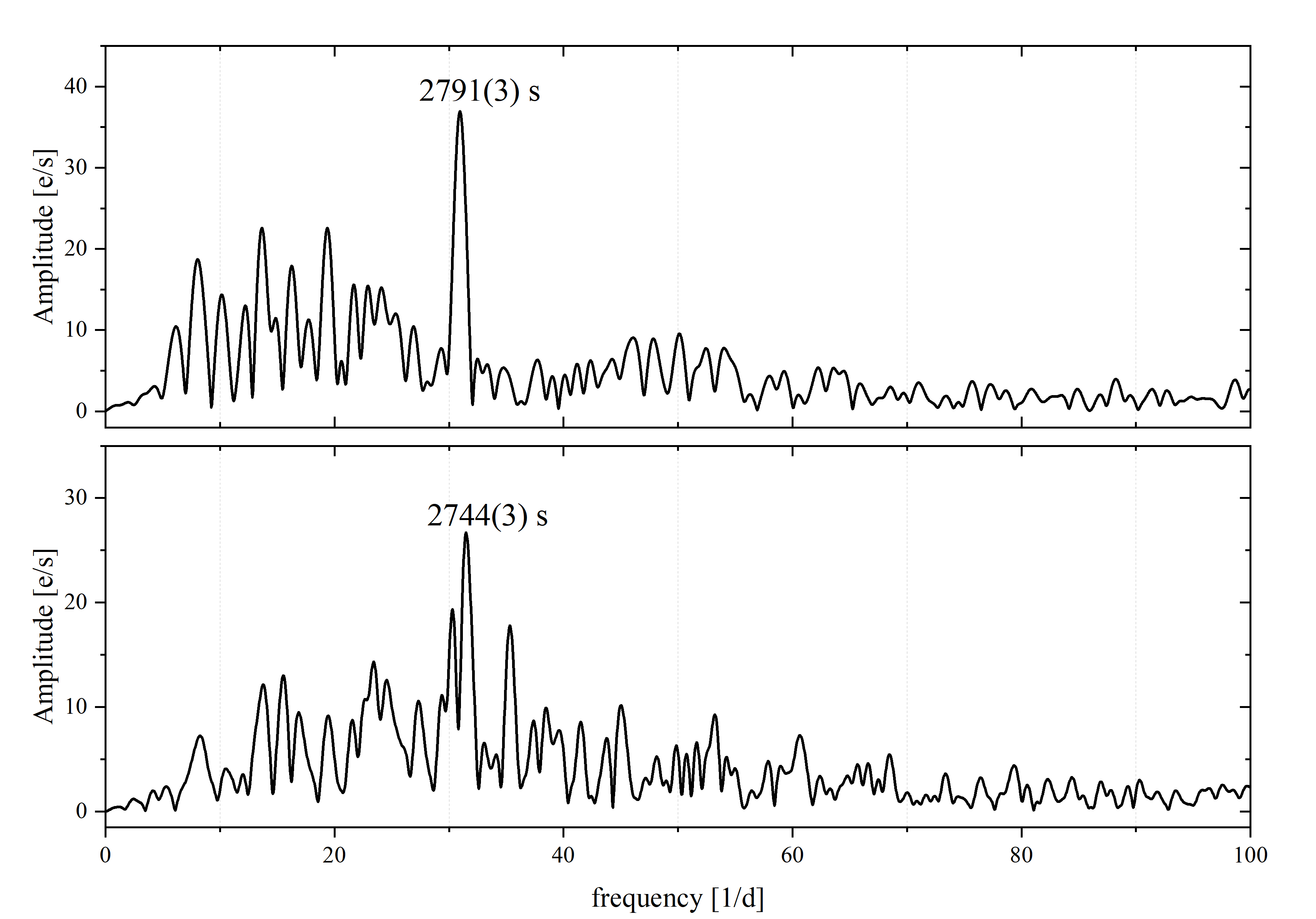}
\caption{Frequency - Amplitude plots for the analysis of QPOs. The top and bottom panels analyze the top detrended light curves for c25 and c52 long outbursts, respectively.\label{fig:QPOp}}
\end{figure}

Based on \textit{TESS} photometry, we found $\sim $ 2160 s QPOs at the top of the long outburst of another Z Cam-type DN (HS 2325+8205; \citealp{Sun2023MNRAS}). Similarly, we found QPOs at the top of the long outburst of AH Her (see Fig. \ref{fig:abc}). We selected the top of the two long outbursts and first removed the trend using a LOESS fit  (see Figs. \ref{fig:QPO1} and \ref{fig:QPO2}) and then analyzed the residuals. The results indicate the presence of QPOs with periods of 2791(3) s (30.95(4) d $ ^{-1} $) and 2744(3) s (31.48(3) d $ ^{-1} $) at the top of the long outbursts of s25 and s52, respectively (see Fig. \ref{fig:QPOp} and Table \ref{tab:freq}). Again we use the least-squares sinusoidal fit of Eq. \ref{eq:sine} for verification  (see Figs. \ref{fig:QPO1} and \ref{fig:QPO2}).
The best-fit results are shown in Table \ref{tab:sine}, with errors in each fitted parameter derived from the covariance. The best-fit periods were 2792(3) s (0.03231(4) d) and 2745(3) s (0.03177(3) d) in agreement with the 2791(3) s and 2744(3) s.

\citet{warner2004rapid} classified QPOs into three categories -- DNO-related, IP-related, and other. With periods of thousands of seconds, the QPOs in AH Her belong to the ``other" class of QPOs.
QPOs are thought to be present in CVs with a high mass transfer rate. Z Cam stars have high mass transfer rates compared to other dwarf novae \citep{dubus2018testing}, so the presence of QPOs in AH Her is reasonable. \cite{Sun2023MNRAS} also found $\sim $ 2160 s QPOs at the top of the long outburst of HS 2325+8205 based on \textit{TESS} photometry, which suggests that the presence of QPOs at the top of the long outburst of Z Cam-type DNe may be a common phenomenon, which further demonstrates that QPOs are associated with high mass transfer rate.

\section{DISCUSSION} \label{sec:DISCUSSION}

\subsection{Outbursts with negative superhumps} \label{sec:DISCUSSION1}

NSHs have been found during regular outbursts, superoutbursts, and quiescence in SU UMa-type DNe, e.g., V503 Cyg \citep{harvey1995superhumps}, ER UMa \citep{Ohshima2012PASJ}, BK Lyn \citep{Kato2013PASJ}, V1504 Cyg \citep{Osaki2013a,Ohshima2014PASJ}, V344 Lyr \citep{Wood2011ApJ}, SDSS J210014.12+004446.0 \citep{Olech2009MNRAS}, and NY Her \citep{Pavlenko2021Ap}. However, there is no uniformity in the period variation of NSHs, with some NSHs becoming longer and some shorter during the outbursts \citep{Smak2013}.
The amplitude of NSHs was found to be increased during the SU UMa outbursts (see Figs. 15 and 20 of \cite{Wood2011ApJ}, and Figs. 4 and 5 of \cite{Osaki2013b}).

\cite{Montgomery2009MNRAS} suggested that the change in NSHs amplitude of a long timescale variation like V1504 Cyg may be related to the change in the tilt angle of the accretion disk.
\cite{Wood2011ApJ} suggested that the period variation of NSHs comes from the distribution of the accretion disc material, and the increase in the mass of the inner disc leads to an increase in the precession period of the accretion disk and an increase in the NSHs period. 
\cite{Osaki2013a,Osaki2013b} suggested that the variation of the NSHs period comes from the variation of the radius of the accretion disk, while the variation of the NSHs amplitude is caused by the addition of the disk component during the outburst in addition to the gas stream component. 
However, \cite{Smak2013} disagreed with \cite{Osaki2013a}'s view on the variation of the radius of the accretion disk, arguing that the periodic variation of the NSHs is complex and the conclusion based on a single star is not representative, and in addition the larger the accretion disk, the smaller the amplitude of the NSHs, while suggesting that the variation of the mass transfer rate may be a cause.

Other types of DNe have also been found to have NSHs.
For example, \cite{ramsay2017v729} and \cite{court2020ex} found NSHs only in the quiescence of Z Cam-type DNe V729 Sgr and EX Dra based on abundant \textit{TESS} outburst data, and no NSHs were found during the outburst. \cite{KimuraKIC94066522020PASJ} found the presence of NSHs in Z Cam-type DNe with IW And-type phenomenon, and the NSHs period varies with the IW And-type phenomenon: the period increases during the brightening and gradually decreases during the quasi-standstill, but the amplitude of the NSHs does not vary significantly. 

Our work is different from the above; we found that the NSHs amplitude is maximum in the quiescence, decreases with the rise of the outburst, is undetectable at the top of the outburst, decreases after the NSHs bounces again in the plateau region, and then enhances again with the fall of the outburst. The most peculiar of these changes comes from the plateau, where there was a noticeable rebound followed by a weakening. Based on the O-C analysis of the NSHs maxima, we find that the period of the NSHs is slightly increased during the outburst except for the top.

The DIM model is widely used to explain normal DNe outbursts, assuming that the mass transfer from the secondary star is relatively stable and that the outburst evolves from an accretion disk thermal viscous instability cycle \citep{lasota2001disc}. It is widely believed that the origin of NSHs is the interaction of the reverse precession of the nodal line in the tilted disk with the binary star motion, in which the accretion stream plays an important role (e.g., \citealp{Barrett1988MNRAS}; \citealp{Patterson1997PASP}; \citealp{Wood2007ApJ...661.1042W}). 
The variations in the amplitude and period leading to the NSHs may arise from changes in the physical and geometric position of the tilted disk, assuming stable mass transfer from the secondary star. 

Studies have shown that DNe accretion disks are larger in outburst than during quiescence \citep{warner1995cat}. 
If the radius of the accretion disk increases during the outburst in AH Her, it will lead to a smaller gravitational potential energy released by the accretion stream \citep{Smak2013}, which is shown as the maximum NSHs amplitude in the quiescence, and the NSHs amplitude decreases as the outburst rises. The variation of the radius of the accretion disk can explain the phenomenon of NSHs amplitude variation except for the plateau region in AH Her. The NSHs at the top of the outburst are undetectable, and it may not be that the NSHs disappear but weaken. The rebound and weakening of the NSHs in the plateau region are unlikely to be caused by the shrinking and expanding of the short-term timescale of the accretion disk because there would be a larger-scale optical variation. 
Therefore, the relationship between NSHs and outbursts is complex but of high research value, and we suggest that the relationship between NSHs and outbursts can be used as a window to investigate the accretion disk model and the origin of NSHs.

\subsection{Tilted disk?} \label{sec:disk}

The NSHs are caused by the reverse precession of the nodal line in a tilted disk with the binary motion. There is a relationship between the precession period, NSHs, and orbital period in beats:
\begin{equation} \label{eq:pso}
\dfrac{1}{P_{\rm prec}}  =\dfrac{1}{P_{\rm nsh}} - \dfrac{1}{P_{\rm orb}}  
\end{equation} 
We determine the NSHs of AH Her as $ P_{\rm nsh} $ = 0.24497(1), so the theoretical precession period is $ P_{\rm prec} $ = 4.77(1) d (0.2098(6) d$ ^{-1} $).

Although no evidence for the tilted disk precession was found in the light curve of AH Her, we found signals that may be related to precession when we performed amplitude, O-C, and Slope analysis of NSHs. 
The study of the amplitude variation found a periodic variation of NSHs amplitude with a period of 2.41(5) d in addition to the variation with the outburst, which is about half of the theoretical precession period of the tilted disk, and was verified in the sinusoidal fit.
In Section \ref{sec:O-C}, we found the O-C of the NSHs, the S$ _{\rm r}$, and the S$ _{\rm d} $ exist with a periodic variation of 2.33(2) d to 2.68(5) d, which is about half of the period of tilted disk precession. Therefore, we suggest that the variation of amplitude, O-C, and the Slope of NSHs may be related to the tilted disk precession.

\section{CONCLUSIONS} \label{sec:CONCLUSIONS}

In this paper, we investigated the relationship between negative superhump and quasi-periodic oscillation and outburst of AH Her based on \textit{TESS} photometry, which is summarized as follows:

Based on the continuous wavelet transform and \texttt{Period04}, we found negative superhumps in AH Her for the first time, with period 0.24497(1) d ($\epsilon ^- \sim 0.051$). 
Combining the variation of the amplitude, the light curves, and the two-dimensional spectrum of the continuous wavelet transform, we found that the negative superhumps amplitude of AH Her varies with the outbursts. The amplitude is the largest in the quiescence and gradually decreases as the outburst rises; when reaching the top of the outburst, negative superhumps are undetectable. The negative superhumps bounce and then decrease in the plateau, and the amplitude gradually increases again as the outburst falls. The variation of amplitude can be explained by the variation of the accretion disk except for the plateau, so the relationship between negative superhumps and outburst is complex but of high research value. We suggest that the relationship between negative superhumps and outbursts can be used as a window to study the accretion disk model and the origin of negative superhumps.

In addition to the variation of the negative superhumps amplitude with the outburst, we found a periodic variation of 2.41(5) d and verified in the sinusoidal fit, which is about half the theoretical period of the tilted disk precession.
The O-C analysis of the negative superhumps maxima indicates that there may be periodic variations in the maxima, but they are not statistically significant. We calculated the rising and declining slopes of negative superhumps to complement the lack of O-C as a proxy study. The results show that the O-C of maxima, as well as the Slope, exists with a period variation ranging from 2.33(2) d to 2.68(5) d, which is about half of the theoretical period of tilted disk precession, so we suggest that the changes in NSHs amplitude, O-C, and slope may be related to the tilted disk precession.

Finally, we found for the first time QPOs with a period of about 2800 s at the top of the long outburst of AH Her, which was verified in a sinusoidal fit. This is similar to \cite{sun2022study}'s discovery of QPOs during the long outburst of another Z Cam-type dwarf nova HS 2325+8205 based on \textit{TESS} photometry. This further suggests that the presence of QPOs during long outbursts of Z Cam-type dwarf novae may be common, which is another example of the origin of QPOs related to the high mass transfer rate. 

\section*{Acknowledgements}

This work was supported by the National Natural Science Foundation of China (Nos. 11933008). We thank all the \textit{TESS} staff for making it possible to study the relationship between negative superhumps, quasi-periodic oscillations, and outbursts in AH Her. We are particularly grateful to our peer reviewers for their professional and detailed advice that has made this work more complete.

		
		\bibliography{sample631}{}
		\bibliographystyle{aasjournal}

	
	\section{Appendix: Supplemental Materials}\label{sec:Appendix}

	\renewcommand\thetable{S1}
	\setlength{\tabcolsep}{2pt}
	\begin{longtable}{ccccccccccccccc}

		\caption{ Parameters of the negative superhumps in AH Her.}\label{tab:S1}\\

		
		\toprule
	maximum-BJD&errors&maximum-flux&errors&E&O-C&Amplitude&errors&Rising Slope&errors&Declining Slope&errors&Sectors\\
	
	[BJD-2457000]&[d]&[e/s]&[e/s]&[n]&[d]&[e/s]&[e/s]&[e/(s$\cdot$d)]&[e/(s$\cdot$d)]&[e/(s$\cdot$d)]&[e/(s$\cdot$d)]&\\
		\midrule
		\endfirsthead
		\multicolumn{13}{c}
		{{\bfseries{Table \thetable\ }continued from previous page}} \\
		\toprule
	maximum-BJD&errors&maximum-flux&errors&E&O-C&Amplitude&errors&Rising Slope&errors&Declining Slope&errors&Sectors\\

	[BJD-2457000]&[d]&[e/s]&[e/s]&[n]&[d]&[e/s]&[e/s]&[e/(s$\cdot$d)]&[e/(s$\cdot$d)]&[e/(s$\cdot$d)]&[e/(s$\cdot$d)]&\\
		\midrule
		\endhead
		\midrule
		\multicolumn{13}{c}
		{{ }} \\
		\endfoot
		\endlastfoot
		1984.6715&0.0052&47.48&31.49&0&0.0000&-&-&-&-&1895.54&770.32&s25\\
		1984.9187&0.0018&101.18&22.52&1&0.0022&222.19&45.82&1468.33&348.88&2543.07&574.25&s25\\
		1985.1399&0.0031&88.47&28.32&2&-0.0215&232.30&52.00&1496.27&395.12&2101.57&532.75&s25\\
		1985.4085&0.0025&-7.53&25.35&3&0.0021&147.79&53.24&1141.76&400.82&1188.67&533.80&s25\\
		1985.6587&0.0023&160.23&23.87&4&0.0073&281.00&52.57&2127.73&434.82&2119.37&487.59&s25\\
		1985.8973&0.0043&137.47&26.17&5&0.0009&245.32&52.46&2108.41&586.92&1927.81&459.59&s25\\
		1986.1449&0.0035&194.38&26.28&6&0.0036&301.22&50.29&2598.24&518.96&2111.95&438.71&s25\\
		1986.4035&0.0013&111.59&20.89&7&0.0173&202.99&43.68&1794.78&425.12&1382.90&326.87&s25\\
		1986.6460&0.0008&155.57&23.86&8&0.0148&250.68&45.64&2305.46&488.98&1646.16&330.52&s25\\
		1986.8923&0.0028&142.37&23.95&9&0.0162&228.52&45.88&2926.54&744.94&1251.43&328.27&s25\\
		1987.1627&0.0017&187.99&26.56&10&0.0416&235.32&52.17&2377.10&586.00&2344.45&730.90&s25\\
		1987.3992&0.0009&133.85&25.29&11&0.0331&195.19&52.44&1153.01&420.95&2213.70&525.90&s25\\
		1987.6241&0.0017&39.35&19.45&12&0.0130&120.08&47.74&1064.58&377.40&889.71&468.41&s25\\
		1987.7953&0.0013&83.61&21.62&13&-0.0607&160.83&50.47&3164.50&1501.08&843.27&247.58&s25\\
		1988.0545&0.0012&51.27&20.24&14&-0.0465&158.38&47.86&2273.25&872.09&1067.31&314.23&s25\\
		1988.3235&0.0010&43.44&19.23&15&-0.0225&124.76&47.25&1740.49&541.46&692.85&441.36&s25\\
		1988.5707&0.0037&57.53&17.63&16&-0.0203&125.37&40.90&661.05&355.71&1621.23&464.37&s25\\
		1988.8149&0.0043&80.78&14.91&17&-0.0211&177.16&38.28&1262.52&290.51&1737.76&519.60&s25\\
		1989.0544&0.0032&43.96&24.09&18&-0.0265&158.18&48.19&948.82&387.42&1605.64&463.07&s25\\
		1989.3084&0.0048&63.19&24.17&19&-0.0175&204.24&43.74&1435.53&385.72&2030.85&519.30&s25\\
		1989.5252&0.0016&74.24&15.73&20&-0.0456&192.72&31.02&1895.38&328.19&1147.87&220.59&s25\\
		1989.7668&0.0013&81.81&16.01&21&-0.0491&173.13&28.69&1831.45&353.55&1106.89&196.80&s25\\
		1990.0194&0.0017&63.87&11.38&22&-0.0414&149.06&24.61&1563.29&279.83&962.44&198.93&s25\\
		1990.2750&0.0009&73.64&11.89&23&-0.0307&158.15&26.58&1471.59&298.56&1095.79&196.98&s25\\
		1990.5180&0.0014&52.67&11.68&24&-0.0327&126.69&25.71&1473.49&318.62&829.95&200.86&s25\\
		1990.7758&0.0013&71.14&12.07&25&-0.0199&134.90&26.73&1075.75&231.50&1119.11&253.41&s25\\
		1991.0283&0.0008&86.18&10.65&26&-0.0123&156.59&24.32&1194.31&230.50&2053.93&347.00&s25\\
		1991.2731&0.0010&102.22&15.21&27&-0.0125&190.55&36.90&1050.48&174.78&2581.10&682.53&s25\\
		1991.5344&0.0009&93.46&12.78&28&0.0038&197.77&36.06&1083.19&262.60&2240.20&350.62&s25\\
		1991.7839&0.0008&87.57&9.90&29&0.0083&196.31&25.44&1199.50&163.32&2465.78&380.24&s25\\
		1992.0153&0.0008&84.13&14.90&30&-0.0052&197.66&30.88&1305.57&225.20&1953.22&325.95&s25\\
		1992.2564&0.0010&49.09&13.09&31&-0.0091&162.60&28.23&1177.85&218.51&1387.92&261.76&s25\\
		1992.4872&0.0013&87.79&14.35&32&-0.0233&206.64&27.30&1744.27&285.38&1469.86&194.07&s25\\
		1992.7537&0.0011&129.95&16.18&33&-0.0018&247.58&28.97&2105.46&255.90&1729.48&250.54&s25\\
		1992.9946&0.0009&162.56&12.57&34&-0.0059&263.89&28.67&2661.90&317.99&1842.79&240.94&s25\\
		1993.2428&0.0005&210.48&11.93&35&-0.0026&293.38&30.41&2768.60&303.10&2621.03&338.21&s25\\
		1993.4885&0.0007&172.11&16.70&36&-0.0019&249.35&34.53&1780.61&290.90&3507.30&606.12&s25\\
		1993.7366&0.0007&162.08&15.24&37&0.0013&239.62&32.98&1386.06&204.04&3060.12&542.26&s25\\
		1993.9785&0.0005&136.27&9.72&38&-0.0018&240.23&25.59&1272.13&191.99&2912.00&274.55&s25\\
		1994.2293&0.0006&120.03&12.13&39&0.0041&254.04&25.27&1611.68&168.86&2734.24&309.44&s25\\
		1994.4771&0.0010&109.94&11.53&40&0.0069&240.34&26.27&1571.42&177.46&2526.99&326.56&s25\\
		1994.7006&0.0008&92.92&10.57&41&-0.0146&217.32&25.51&1697.25&218.57&1664.95&215.68&s25\\
		1994.9338&0.0013&113.61&13.49&42&-0.0264&224.63&26.68&2255.15&314.02&1432.73&194.86&s25\\
		1995.1929&0.0009&130.83&12.06&43&-0.0122&219.82&24.32&2106.79&260.58&1383.51&185.46&s25\\
		1995.4344&0.0011&137.20&11.91&44&-0.0157&204.39&23.44&2363.30&338.23&1366.68&183.58&s25\\
		1997.1610&0.0009&121.61&12.07&51&-0.0038&247.85&26.00&1845.75&210.14&2033.48&251.50&s25\\
		1997.4020&0.0007&138.92&14.47&52&-0.0079&260.98&28.15&2148.06&266.54&1973.24&230.52&s25\\
		1997.6457&0.0012&135.54&19.53&53&-0.0091&238.64&34.85&2318.78&334.58&1520.05&293.42&s25\\
		1997.8904&0.0009&172.18&14.83&54&-0.0094&257.50&34.33&2549.67&401.24&2059.74&325.67&s25\\
		1998.1406&0.0006&198.76&14.51&55&-0.0042&287.45&35.35&2294.79&326.00&2312.61&316.11&s25\\
		1998.3876&0.0006&178.86&14.30&56&-0.0022&290.94&32.64&2209.93&320.67&4440.34&592.06&s25\\
		1998.6374&0.0005&189.14&13.52&57&0.0027&309.70&32.44&1798.63&191.09&3759.70&529.74&s25\\
		1998.8876&0.0005&208.85&13.28&58&0.0080&322.47&34.01&1846.31&221.40&3455.98&405.22&s25\\
		1999.1351&0.0007&149.81&15.02&59&0.0105&279.87&33.36&1768.82&249.63&3074.79&381.95&s25\\
		1999.3842&0.0005&137.21&11.69&60&0.0146&285.48&26.79&1793.62&198.39&3206.86&345.02&s25\\
		1999.6326&0.0007&124.23&12.16&61&0.0181&278.03&25.89&1787.51&188.52&2934.25&346.24&s25\\
		1999.8628&0.0009&159.44&14.74&62&0.0032&300.63&28.18&2281.87&249.56&2342.22&258.54&s25\\
		2000.1025&0.0008&185.58&14.17&63&-0.0020&313.73&28.30&2736.20&277.16&2274.38&250.18&s25\\
		2000.3542&0.0008&182.58&18.91&64&0.0047&302.69&33.75&2675.56&344.69&2073.74&265.66&s25\\
		2000.5982&0.0006&202.62&14.39&65&0.0038&315.28&28.70&3158.94&351.97&3123.94&329.06&s25\\
		2000.8442&0.0005&234.90&14.98&66&0.0047&345.79&28.74&2369.24&222.35&3488.65&330.64&s25\\
		2001.0935&0.0004&239.60&13.57&67&0.0091&354.76&27.97&2338.75&198.21&4075.19&390.77&s25\\
		2001.3463&0.0004&222.55&13.96&68&0.0169&352.42&33.37&2071.78&195.10&3820.70&453.67&s25\\
		2001.5935&0.0004&197.97&13.29&69&0.0192&345.45&34.49&2227.41&263.33&3652.57&383.12&s25\\
		2001.8403&0.0005&172.59&13.33&70&0.0210&322.56&28.10&2173.10&234.54&3655.61&330.31&s25\\
		2002.0863&0.0005&166.85&13.05&71&0.0220&316.25&24.69&1967.44&164.70&3166.03&302.85&s25\\
		2002.3260&0.0006&185.93&13.93&72&0.0167&334.85&26.64&2444.15&220.86&2762.18&253.38&s25\\
		2002.5690&0.0008&190.44&17.45&73&0.0148&334.69&31.10&2719.02&279.41&2676.47&289.90&s25\\
		2002.8135&0.0009&180.95&19.26&74&0.0143&318.81&36.87&2712.88&320.62&2278.28&316.55&s25\\
		2003.0607&0.0007&223.97&18.88&75&0.0166&350.07&39.90&3241.36&399.03&2347.10&300.45&s25\\
		2003.3080&0.0004&225.57&12.45&76&0.0189&347.19&34.62&3435.08&380.78&3092.37&346.19&s25\\
		2003.5555&0.0006&241.19&19.67&77&0.0214&365.44&45.28&2702.60&343.77&3190.40&466.21&s25\\
		2003.8055&0.0004&208.71&11.94&78&0.0264&340.21&42.21&2473.84&327.37&4148.52&602.83&s25\\
		2004.0573&0.0006&184.11&14.46&79&0.0333&340.59&49.07&1910.80&295.70&4386.07&722.93&s25\\
		2004.3045&0.0007&94.79&11.32&80&0.0355&296.82&40.72&1635.70&311.19&4031.08&473.28&s25\\
		2004.5492&0.0019&46.82&16.70&81&0.0352&245.82&35.20&1675.32&258.64&2425.66&426.40&s25\\
		2004.7899&0.0029&108.17&16.03&82&0.0309&244.33&30.94&1830.17&248.46&1958.83&345.49&s25\\
		2005.0037&0.0053&109.51&11.78&83&-0.0002&223.22&24.47&2004.00&358.09&1597.15&218.24&s25\\
		2005.2410&0.0014&87.70&10.98&84&-0.0078&208.44&26.60&2334.24&289.44&1248.97&209.56&s25\\
		2005.4938&0.0008&152.16&10.53&85&0.0000&261.57&30.59&3030.79&411.58&1514.75&199.90&s25\\
		2005.7447&0.0008&141.91&10.69&86&0.0059&249.65&29.03&2925.36&438.91&1557.94&186.83&s25\\
		2006.0028&0.0019&128.02&10.74&87&0.0190&242.49&27.23&2573.29&378.33&1625.14&213.48&s25\\
		2006.2617&0.0006&125.89&11.34&88&0.0330&229.40&29.80&2206.42&288.68&1746.87&290.81&s25\\
		2006.5229&0.0015&88.55&12.03&89&0.0492&167.98&32.92&1321.18&271.72&2125.80&532.17&s25\\
		2006.7748&0.0009&85.35&10.60&90&0.0561&164.48&28.35&852.52&187.36&2015.96&358.17&s25\\
		2007.0236&0.0009&16.50&12.55&91&0.0600&142.33&26.43&668.33&180.53&1932.28&332.22&s25\\
		2007.2806&0.0013&57.67&13.92&92&0.0720&144.18&26.99&1317.96&186.20&1008.00&421.77&s25\\
		2007.5380&0.0012&196.86&12.00&93&0.0845&155.91&27.77&1116.07&147.48&1389.83&474.14&s25\\
		2008.4480&0.0012&197.21&16.32&97&0.0146&-&-&-&-&1420.20&322.49&s25\\
		2008.6966&0.0019&181.38&19.49&98&0.0182&244.73&41.98&2191.81&609.44&1753.08&247.90&s25\\
		2008.9389&0.0009&82.69&16.61&99&0.0155&180.85&41.07&2331.80&672.69&1341.59&279.21&s25\\
		2009.1898&0.0010&89.62&14.83&100&0.0214&-&-&1520.52&300.27&-&-&s25\\
		2718.7841&0.0009&115.93&15.08&0&0.0000&167.74&55.17&1700.00&708.27&2061.18&681.28&s52\\
		2719.0359&0.0013&133.87&19.44&1&0.0069&192.38&49.93&1270.00&388.41&1841.58&511.44&s52\\
		2719.2836&0.0015&99.35&19.37&2&0.0096&177.06&45.37&970.55&316.07&2306.75&595.42&s52\\
		2719.5255&0.0061&19.27&17.09&3&0.0065&124.61&41.58&848.81&336.20&1174.01&463.41&s52\\
		2719.7540&0.0060&55.57&21.94&4&-0.0099&136.87&43.67&1248.14&420.45&917.30&395.70&s52\\
		2719.9882&0.0031&77.50&15.91&5&-0.0207&155.28&38.14&1296.89&406.09&1295.07&333.82&s52\\
		2720.2418&0.0011&88.40&16.72&6&-0.0121&171.36&41.20&1512.51&354.98&965.08&272.77&s52\\
		2720.4869&0.0011&168.96&23.82&7&-0.0119&-&-&3028.93&723.33&-&-&s52\\
		2724.9121&0.0005&197.45&22.70&25&0.0038&236.36&41.79&1528.34&389.06&2319.53&349.93&s52\\
		2725.1543&0.0032&47.52&21.47&26&0.0011&200.87&40.55&1414.99&431.93&1705.53&322.84&s52\\
		2725.3929&0.0011&-26.18&15.20&27&-0.0053&198.44&33.65&1818.94&389.68&1407.44&268.18&s52\\
		2725.6606&0.0013&-57.19&22.13&28&0.0174&135.55&46.79&1796.87&485.82&463.52&324.76&s52\\
		2725.9107&0.0011&60.12&23.36&29&0.0225&124.42&50.60&2331.14&745.65&471.11&434.19&s52\\
		2726.1417&0.0016&157.58&20.64&30&0.0086&128.84&43.95&1310.82&451.03&898.58&391.45&s52\\
		2726.3913&0.0023&217.35&18.77&31&0.0132&155.97&40.88&1258.23&352.38&1017.46&310.92&s52\\
		2726.6547&0.0013&232.35&23.24&32&0.0316&212.85&49.36&1347.31&410.41&1743.82&387.40&s52\\
		2726.9217&0.0006&159.42&21.89&33&0.0536&222.38&47.14&1660.46&486.51&1651.41&318.13&s52\\
		2727.1678&0.0016&70.18&31.55&34&0.0548&179.56&59.73&1769.25&654.37&1257.44&466.44&s52\\
		2728.8007&0.0016&162.59&48.16&41&-0.0271&-&-&-&-&3076.59&920.31&s52\\
		2729.0505&0.0006&59.66&52.72&42&-0.0223&272.37&92.17&2034.05&835.97&2449.22&737.52&s52\\
		2729.3008&0.0004&88.08&54.83&43&-0.0169&316.15&87.63&2264.43&628.83&2624.62&773.90&s52\\
		2729.5460&0.0016&121.08&52.26&44&-0.0167&351.01&89.58&2934.85&796.92&2973.80&837.45&s52\\
		2729.7978&0.0026&134.20&43.72&45&-0.0098&314.56&86.84&2635.78&664.48&2190.10&822.31&s52\\
		2730.0559&0.0006&179.32&36.32&46&0.0033&338.33&93.08&2353.27&673.41&3083.77&921.03&s52\\
		2732.9645&0.0016&96.25&25.44&58&-0.0278&-&-&-&-&2558.94&720.00&s52\\
		2733.2151&0.0016&126.74&20.38&59&-0.0222&249.56&48.95&1656.03&364.64&2366.50&513.67&s52\\
		2733.4639&0.0031&140.46&23.04&60&-0.0182&-&-&1665.39&362.81&-&-&s52\\
		2737.6347&0.0011&100.96&14.86&77&-0.0120&162.39&36.43&904.01&287.72&2654.88&597.25&s52\\
		2737.8913&0.0006&113.20&12.53&78&-0.0003&215.97&33.70&1177.56&207.72&3064.53&548.82&s52\\
		2738.1390&0.0008&125.39&15.09&79&0.0024&223.14&36.93&1310.79&217.98&3115.05&670.23&s52\\
		2738.3859&0.0010&86.49&16.26&80&0.0043&188.98&35.77&991.61&237.63&2339.11&452.23&s52\\
		2738.6029&0.0018&58.28&18.58&81&-0.0236&167.66&37.18&1325.49&300.75&1522.57&463.05&s52\\
		2738.8131&0.0030&56.19&19.58&82&-0.0584&164.39&40.68&1539.02&503.02&1036.93&283.82&s52\\
		2739.0707&0.0026&88.04&13.69&83&-0.0458&197.98&32.98&2120.88&480.33&1206.44&224.46&s52\\
		2739.3365&0.0024&113.24&19.07&84&-0.0250&212.29&37.08&2118.67&432.82&1290.35&265.78&s52\\
		2739.5776&0.0024&142.35&21.49&85&-0.0288&229.26&44.05&2816.76&604.83&1643.83&408.51&s52\\
		2739.8312&0.0021&157.93&17.12&86&-0.0202&238.81&38.76&2065.31&453.15&1965.29&330.78&s52\\
		2740.0735&0.0014&126.64&22.10&87&-0.0228&195.42&45.27&1686.59&351.72&1930.39&620.53&s52\\
		2740.3454&0.0070&115.99&18.94&88&0.0041&210.89&45.50&993.13&331.51&4162.70&1516.97&s52\\
		2740.5752&0.0029&125.53&24.29&89&-0.0111&247.32&46.20&1506.10&344.19&2455.70&566.22&s52\\
		2740.8228&0.0033&124.50&22.32&90&-0.0084&258.95&43.84&1571.84&331.43&2946.76&601.84&s52\\
		2741.0650&0.0021&113.63&19.91&91&-0.0112&271.18&40.60&1839.37&326.14&2612.75&450.23&s52\\
		2741.2937&0.0006&141.89&20.55&92&-0.0275&283.19&39.87&2414.46&348.51&1979.96&350.13&s52\\
		2741.5517&0.0069&134.39&27.21&93&-0.0145&241.83&47.01&2087.78&542.58&1939.23&604.58&s52\\
		2741.7839&0.0017&159.76&19.33&94&-0.0272&259.85&43.38&2159.96&464.65&2214.05&441.53&s52\\
		2742.0384&0.0018&149.74&30.09&95&-0.0177&254.47&59.87&1948.39&479.71&2081.80&565.91&s52\\
		2742.2879&0.0014&137.24&16.76&96&-0.0132&266.31&48.43&1833.93&410.76&2984.40&567.43&s52\\
		2742.5355&0.0015&143.05&18.93&97&-0.0105&307.34&47.60&2023.81&377.08&3730.98&629.68&s52\\
		2742.7815&0.0005&207.89&21.52&98&-0.0095&356.26&52.65&2337.83&309.07&4299.07&847.62&s52\\
		2743.0358&0.0023&200.96&23.57&99&-0.0002&-&-&1852.17&378.61&-&-&s52\\
		\bottomrule
	\end{longtable}

		
		
	\end{document}